\documentclass[journal,twocolumn]{IEEEtran}
\usepackage{xcolor}
\usepackage{mathtools}
\usepackage{epsfig,makeidx,color,subfigure}
\usepackage{amsmath,amssymb,bbm,enumitem}
\usepackage{cite,graphicx,lipsum}
\usepackage[switch,pagewise]{lineno}
\usepackage{hyperref}
\hypersetup{
        colorlinks = true,
        citecolor=blue,
}
\usepackage{draftfigure}

\def\be{ \begin{equation} }
\def\ee{ \end{equation} }
\def\bea{ \begin{eqnarray} }
\def\eea{ \end{eqnarray} }

\def\b0{{\bf 0}}

\ifCLASSOPTIONonecolumn
\else

\fi

\usepackage{geometry}
\begin{document}

\title{Enabling the Wireless Metaverse via\\ Semantic Multiverse Communication}

\author{
Jihong Park, Jinho Choi, Seong-Lyun Kim, and Mehdi Bennis\thanks{
J. Park and J. Choi are with the School of Information Technology, Deakin University. VIC 3220, Australia (email: \{jihong.park, \ jinho.choi\}@deakin.edu.au). 

S.-L. Kim is with the School of Electrical and Electronic Engineering, Yonsei University, Seoul 03722, Korea (email: slkim@yonsei.ac.kr). 

M. Bennis is with the Centre for Wireless Communications, University of Oulu, Oulu 90014, Finland (email: mehdi.bennis@oulu.fi).

This research was supported by the Australian Government through the Australian Research
Council's Discovery Projects funding scheme (DP200100391).}}

\maketitle
\begin{abstract}
Metaverse over wireless networks is an emerging use case of the sixth generation (6G) wireless systems, posing unprecedented challenges in terms of its multi-modal data transmissions with stringent latency and reliability requirements. Towards enabling this wireless metaverse, in this article we propose a novel semantic communication (SC) framework by decomposing the metaverse into human/machine agent-specific \emph{semantic multiverses (SMs)}. An SM stored at each agent comprises a semantic encoder and a generator, leveraging recent advances in generative artificial intelligence (AI). To improve communication efficiency, the encoder learns the \emph{semantic representations (SRs)} of multi-modal data, while the generator learns how to manipulate them for locally rendering scenes and interactions in the metaverse. Since these learned SMs are biased towards local environments, their success hinges on synchronizing heterogeneous SMs in the background while communicating SRs in the foreground, turning the wireless metaverse problem into the problem of \emph{semantic multiverse communication (SMC)}. Based on this SMC architecture, we propose several promising algorithmic and analytic tools for modeling and designing SMC, ranging from distributed learning and multi-agent reinforcement learning (MARL) to signaling games and symbolic AI.

\end{abstract}


\ifCLASSOPTIONonecolumn
\baselineskip 28pt
\fi


\section{Introduction} \label{Sec:Intro}

Supporting the metaverse has emerged as a high-stake application in the sixth generation (6G) wireless systems~\cite{hashash2022towards}. Such \emph{wireless metaverse} is envisaged to be sharply distinguished from the current primitive metaverse applications, in terms of its fidelity, interactivity, and mobility. Departing from the current metaverse -- such as Roblox or Second Life that is nothing but a third-person massive multiplayer online (MMO) game, the upcoming wireless metaverse will use all human perceptions to deliver high-fidelity immersive experiences from a first-person perspective.
These experiences will not only be limited within human agents and equivalently their virtual avatars, but will be extended boundlessly to interactive virtual objects that can be associated with intelligent machine agents in the physical space. What is more, such extended reality (XR) experiences will be offered to mobile agents in a wide physical area, spurred by the upcoming self-driving cars and their spaces on wheels. 


In short, enabling the wireless metaverse boils down to the problem of supporting (i) realistic multi-modal interactions among (ii) a variety of human and machine agents in (iii) wide-range mobile scenarios. Existing approaches may address these issues one by one, but certainly not all of them simultaneously. For instance, large bandwidth in the millimeter-wave or terahertz spectrum enables transferring large volumes of 360-degree videos in enhanced mobile broadband (eMBB) communication \cite{quamcomm:XR}, yet its required sharp beams are ill-suited for mobile scenarios. Meanwhile, prioritized resource allocation in ultra-reliable and low-latency communication (URLLC) can enable latency-sensitive interactions such as haptic feedback, but its effectiveness is negated under multimodal interactions or equivalently a mixture of synchronous URLLC-eMBB traffic \cite{8647208}. Lastly, limitless virtual interactions result in numerous use cases and events, and it is practically impossible to manually identify and respond to every single vertical's physical communication requirements.



\begin{figure}[t]
\centering
\includegraphics[width=\columnwidth]{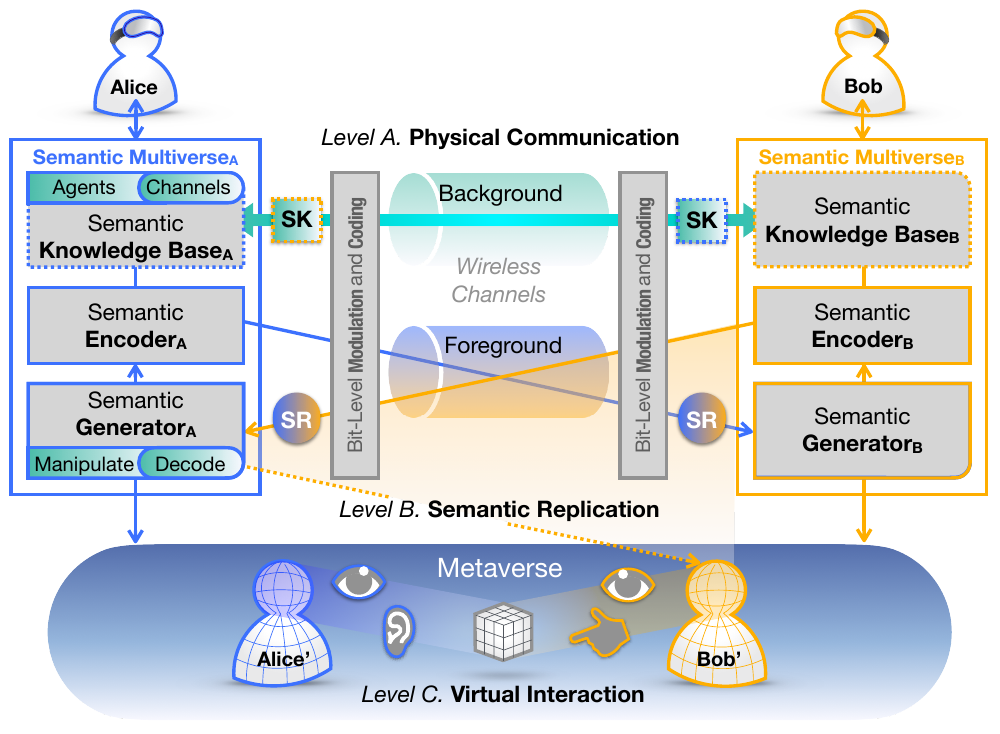}  
\caption{\emph{Semantic multiverse communication (SMC)} between Alice and Bob running their local \emph{semantic multiverses (SMs)} for distributed metaverse rendering. Alice and Bob communicate the \emph{semantic representations (SRs)} of their multi-modal data in the foreground and \emph{semantic knowledges (SKs)} for synchronous SMs in the background.
} \label{Fig:1} 
\end{figure}

To resolve such challenging wireless metaverse problems, as illustrated in Fig.~\ref{Fig:1}, in this article we put forward a new perspective for decomposing the wireless metaverse into agent-specific \emph{semantic multiverses (SMs)}. Technically, this decomposition can be implemented by distributing the functionalities of a central server running the metaverse into mobile edge computing devices \cite{park2018minimizing}. Conceptually, this idea is inspired from human cognition theory stating that every human understands and reacts to the world through one's sensorimotor system and local semantic memory~\cite{tulving1972episodic} that stores general knowledge abstracted from past experiences. Likewise, an SM is not just a shard of the metaverse in raw data, but stores local-yet-general knowledge of common patterns, long-term relations, and their invariant concepts (e.g., prior distributions, causes, or generative processes of multi-modal raw data). The SM not only stores these \emph{semantic knowledges (SKs)} but is also a tunable function. Each SM can pre/post-process raw data, and manipulate the resultant \emph{semantic representation (SR)} for locally rendering high-fidelity objects and multi-modal feedback in the virtual space, drawing on recent advances in generative artificial intelligence (AI) (e.g., text-to-3D object creation \cite{https://doi.org/10.48550/arxiv.2211.10440} together with object-to-vibrotactile feedback).

Finally, SMs ought to be interoperable to support boundless virtual interactions, warranting inter-SM communication for wireless metaverse. In essence, the wireless metaverse sits the confluence of \emph{physical}, \emph{digital}, and \emph{virtual } worlds \cite{quamcomm:XR}. Meanwhile, according to Shannon and Weaver \cite{ShannonWeaver:1949}, communication fundamentally aims to achieve three-level objectives: a \emph{technical} level of transferring bits; a \emph{semantic} level of delivering the meanings of the bits; and a task-specific or \emph{pragmatic} level of improving the delivered meanings' effectiveness in a given task. By blending and juxtaposing these two viewpoints, we propose the following three-level communication objectives for enabling the wireless metaverse.

\begin{itemize}
\item \textbf{Level A -- Physical Communication}: Agents exchange SRs and SKs through physical communication over wireless channels.

\item \textbf{Level B -- Semantic Replication}: To connect different SMs, agents need to use SRs grounded in multiple SMs as a universal language. Such SRs do not deliver a bit-level description of each frame, but the semantics of intended changes in the virtual world, enabling the digital replication of physical objects with much reduced data payload.

\item \textbf{Level C -- Virtual Interaction}: 
Imbuing SR communication in the foreground, agents should also communicate SKs in the background so as to synchronize their SMs. While each SM is continuously updated by the received SRs, local decision-making, and environmental changes, the SKs therein are slowly updated thanks to their generality, enabling virtual interactions with much relaxed synchronization requirements. 

\end{itemize}


The proposed shift in perspective turns the problem of supporting the wireless metaverse into designing communication with and across locally updated SMs, coined \emph{semantic multiverse communication (SMC)}. SMC is rooted in the emerging paradigm of semantic communication (SC) \cite{9955525,https://doi.org/10.48550/arxiv.2108.05681,Choi22_VTC,CLP_22,Guler18} that aims to effectively deliver the meanings of source symbols for a given task, i.e., semantic and pragmatic levels in \cite{ShannonWeaver:1949}. While existing SC frameworks deal with relatively simple tasks such as raw data reconstruction \cite{9955525}, SMC focuses on a distributed generative task that should guarantee the same semantics for consistent virtual experiences across agents (e.g., visualizing the same semantic scene from different first-person angles) and data modalities (e.g., making a sound and a visual effect consistently in response to touching the same object). In this article, we aim to introduce promising analytic and algorithmic approaches for modeling and designing SMC, through the lens of signaling games, distributed learning, multi-agent reinforcement learning (MARL), and symbolic AI.











\section{Semantic Multiverse Communication Systems} \label{Sec:SMC}

\subsection{AI-Native and Knowledge-Guided SM Architecture} \label{Sec:Architecture}

To achieve the Levels A-C, we put forward an \emph{AI-native yet knowledge-guided SM architecture} as illustrated in Fig.~\ref{Fig:1}. Specifically, each human/machine agent is associated with an SM that provide SRs and SKs defined as follows.
\begin{itemize}
\item \textbf{Semantic Representation (SR)}: It is a learned embedding of high-dimensional multi-modal data describing physical/virtual objects and their interactions. Leveraging deep representation learning, an SR is in a form of the (hidden-layer) activation in a neural network (NN).

\item \textbf{Semantic Knowledge (SK)}: An SK stores the knowledge on the causal/logical structures of SRs,  in a form of a graph (e.g., a KG or a structured causal model), a set of symbolic expressions (e.g., logic programming or mathematical equations), or an NN (e.g., the weight parameters of an NN). An NN-based SK is constructed by training the NN until it can generate/process the SRs. Graphical and symbolic SKs can be constructed via deep learning based or neuro-symbolic knowledge/graph discovery methods. 

\end{itemize}

To produce, manipulate, and store SRs/SKs, each SM comprises three components: a semantic encoder, a semantic decoder, and a semantic knowledge base, as elaborated~next.
\begin{itemize}
\item \textbf{Semantic Encoder}: It is an NN pre-processesing raw data and producing their SRs. To understand SRs between communiating agents, a semantic encoder is jointly trained with its paired semantic generator, which will be discussed through the lens of a game-theoretic modeling in Sec.~\ref{Sec:SignalingGame}. The resultant SRs are biased towards the semantic encoder-generator NNs' training environment such as source/generated data and wireless channels between them. To cope with this problem, SRs can be grounded in multiple training environments or equivalently multiple SMs through physical communication or virtual emulation, to be discussed in Sec.~\ref{Sec:DistributedML}.


\item \textbf{Semantic Generator}: It combines the functionalities of \emph{decoding} the received SRs and \emph{manipulating} them for rendering virtual scenes and interactions in an intended form in terms of modality, angle, fidelity, etc. Spurred by recent advances in generative AI, the architecture can be a generative NN. As a special case, when the intended output form is identical to raw data input, it only performs a decoding functionality, and the resulting architecture together with the semantic encoder is tantamount to the autoencoder (AE) NN.

\item \textbf{Semantic Knowledge Base (KB)}: While semantic encoder/generator takes the role of a tunable function in an SM, a KB takes the role as a memory that stores SKs on its own/other \emph{agents}, as well as inter-agent SKs such as wireless \emph{channels}.  The KB is associated with the semantic encoder/generator, externally or internally as an augmented input or a training regularizer, respectively. By leveraging other/inter-agent SKs in the KB, each SM can locally emulate SMC within its KB.


\end{itemize}

As a special case, when multiple SMs are jointly trained via local emulation with only NN-based SKs (i.e., semantic encoder/generator NNs), each trained SM amounts to a single NN storing other/inter-agent SKs. This \emph{neural SM} will be studied in Sec.~\ref{Sec:DistributedML} and \ref{Sec:Comm-MARL}, which can be transformed into a \emph{symbolic SM} as we will demonstrate in Sec.~\ref{Sec:NS_Transform}.

The functionalities of the aforementioned three SM components are operated before and after the modulation and coding blocks in classical communication system architecture. Implementing these SM components can be either physically separated as depicted in Fig.~\ref{Fig:1}, or jointly combined with a deep learning based architecture as considered in deep learning based joint source and channel coding (DeepJSCC) architectures \cite{9955525}.


It is also worth noting that there is a knowledge-native architecture considered in a language-based application of a multi-agent system (MAS). In this MAS, machine agents autonomously communicate with each other using an agreed language such as knowledge query and manipulation language (KQML). This presumes a complete general knowledge such as an ontology graph (e.g., \emph{Action} \!\!$\overset{\text{causes}}{\xrightarrow{\hspace*{10pt}}}$\!\!
 \emph{Event}). By feeding the actual data, the ontology graph produces a knowledge graph (KG) (e.g., ``\emph{taking a red pill}" \!\!$\overset{\text{causes}}{\xrightarrow{\hspace*{10pt}}}$\!\!
 ``\emph{quitting the metaverse}"), thereby rendering multimodal responses in the virtual space. Such a knowledge-native SM architecture requires pre-programing all possible events and their relations. This becomes too costly in the metaverse that is much more interactive with more freedom under the co-presence of both human and machine agents.

\subsection{Foreground and Background Semantic Communication}

The three key components of SM are tunable and updated through virtual-space interactions or equivalently communicating and processing SRs and SKs in the physical space. This contrasts with classical transceiver designs that are pre-programmed and fixed, calling for the following SMC operations.

\begin{itemize}
    \item \textbf{Foreground Communication}: It refers to the communication between the semantic encoder and the semantic generator, conveying the SRs of multi-modal data for the generator's task.
    
    \item \textbf{Background Communication}: It refers to the communication to ensure rendering consistency across different agents' SMs, by exchanging update messages of the SKs (e.g., output error gradients or a fraction of SKs).
    
\end{itemize}

Foreground and background communications focus on delivering SRs and SKs, respectively, and aim to improve their effectiveness in a distributed metaverse rendering task. 
They operate in a closed-loop in the sense that SKs improve SRs that update SKs.
They also have different time-scales. Unlike long-term SKs, SRs are exchanged in real-time and on a much faster time-scale. In this respect, SMC is a two-way asynchronous SC. As a special case, SMC covers several existing SC applications. For instance, the deep learning based SC (DeepSC) framework \cite{9955525} coincides with foreground communication tasked with raw data reconstruction between two agents and their SKs are concurrently trained NNs. With many agents targetting non-reconstruction tasks, SMC amounts to MARL with communication \cite{https://doi.org/10.48550/arxiv.2108.05681}.

In SMC, the SR-SK relationship can be learned iteratively via interactions. Such a relationship is often causal and logical, rather than statistical as in classical communication systems. In this respect, SMC calls for new analytic and modeling tools as we shall elaborate in the following sections.





\section{Distributed Learning for SMC} \label{Sec:DistributedML}

\subsection{DeepSC Limitations under Heterogeneous SMs} \label{Sec:DeepSC}

DeepSC is an AI-native communication approach where an NN pre/post-processes raw data either separately or jointly with source and channel coding \cite{9955525}. DeepSC is also promising for converting multi-modal metaverse data into SRs. The architecture of DeepSC rests on an AE NN that is decomposed into an encoder NN and a decoder NN. Suppose that Alice and Bob store the encoder and decoder NNs, respectively, and jointly train them though communication, i,e., by Bob's decoding the SR encoded and transmitted by Alice through the channel between them. We can interpret these encoder and decoder NNs as SKs that are dependent on the training data and channel between Alice and Bob. The resultant data-channel specific SMs are interoperable between Alice and Bob, yet entail a compatibility issue with other agents. To illustrate this, consider the SKs of Carol and David, trained in a different data-channel environment. In this case, the SR generated by Alice's SM cannot be decoded by David's SM that is incompatible with Bob's SM, incurring the problem of \emph{heterogeneous SMs}.


\begin{figure}[t]
\centering
\includegraphics[width=.9\columnwidth]{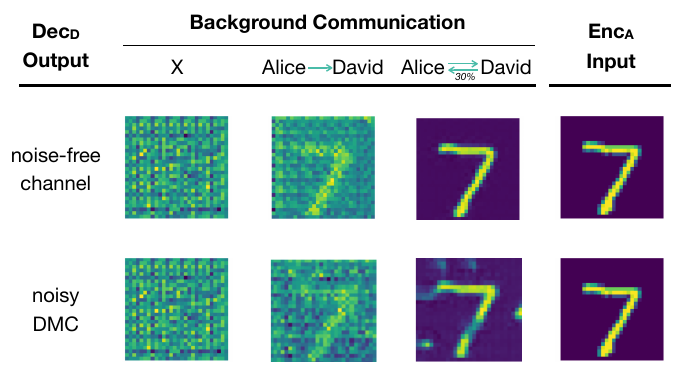} 
\caption{An example of FL and SL for SMC where Alice-Bob DeepSC and Carol-David DeepSC are trained using the MNIST dataset over noise-free channels and the Fashion MNIST dataset over noisy discrete memoryless channel (DMC), with respect to no background communication, only downloading David's generator, and both downloading and partially uploading David's re-trained generator.} \label{Fig:3}
\end{figure}

\subsection{Federated and Split Learning for Heterogeneous DeepSC} \label{Sec:FLSL}
Recalling that SKs are NNs in DeepSC, distributed learning can resolve the heterogneous SM problem or equivalently Alice-David interactions. A na\"ive way is to apply federated learning (FL) that periodically exchanges and averages the Alice-Bob NNs and the Carol-David NN \cite{park2021communication}. Unfortunately, FL is ill-suited in non-independent and identically distributed (non-IID) data or equivalently different data-channel environments in SMC. Split learning (SL) is another straightforward solution that continually trains the Alice-Bob NN, followed by the Alice-David NN \cite{park2021communication}. While SL is less sensitive to non-IID environments, it should exchange every forward activation and its backward gradient throughout the entire training process.


Alternatively, we combine FL and SL to complement their respective drawbacks as follows.
\begin{enumerate}
    \item In the background communication, David sends its trained decoder NN to Alice.
    \item Alice connects its trained encoder NN with the received decoder NN from David, and locally re-train the connected~NN.
    \item To improve background communication efficiency, Alice freezes a part of or the entire decoder NN during the re-training process, followed by sending only the re-trained decoder fraction back to David.
    
    
    \item The re-trained Alice's encoder is now capable of producing SRs that can be better decoded by David.
\end{enumerate}

Fig.~\ref{Fig:3} shows a toy implementation of this scenario where the Alice-Bob NN and the Carol-David NN are trained using the MNIST dataset (numbers images) and the Fashion MNIST (clothes images), respectively. The result shows that even when the decoder NN is entirely frozen, encoder re-training is certainly effective (e.g., for a classification task), but insufficient to achieve high fidelity. To fix this issue, in 3), we consider unfreezing 1/3 of the decoder NN, and sending this re-trained partial decoder NN back to David. This additional background communication achieves higher fidelity reconstruction at David, even under a discrete memoryless channel (DMC) noise.

\begin{figure}[t]
\centering 
\includegraphics[width=\columnwidth]{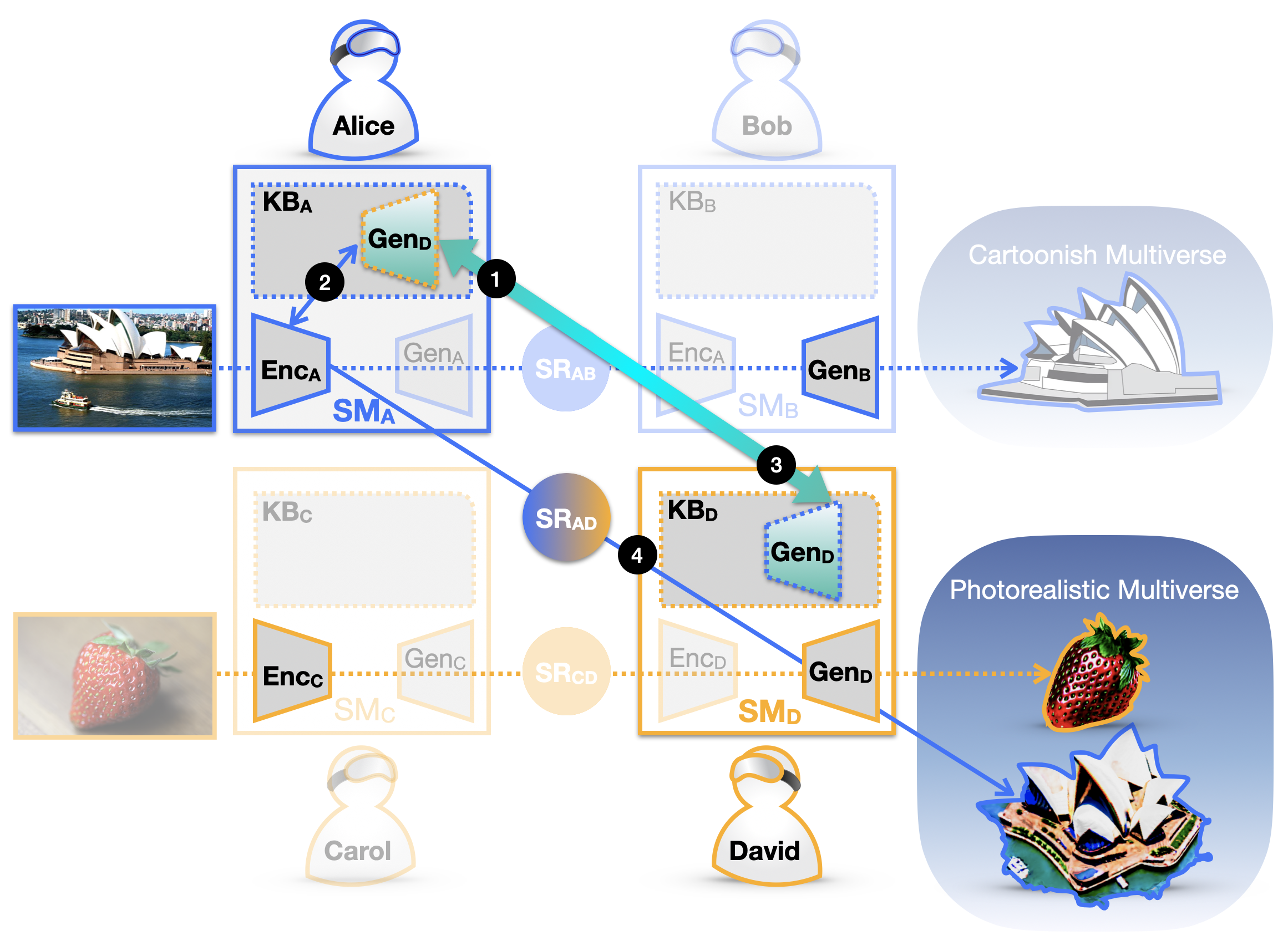} 
\vspace{-18pt}
\caption{The operations of federated learning (FL) and split learning (SL) for SMC where the DeepSC of Alice and Bob are incompatible with that of Carol and David trained over different data and channel environments, which is overcome by communicating David's generator in the background, and re-training Alice's encoder (and a fraction of David's generator) via emulating Alice-David foreground communication within Alice's KB. \protect\footnotemark}
 \label{Fig:2} 
\end{figure}

To extend this FL-SL based framework to SMC, the decoders in the DeepSC architectures can be replaced with semantic generators, as exemplified in Fig.~\ref{Fig:2}. Here, Alice interacts with Bob in a multiverse rendered in a cartoon style, while David does with Carol in a multiverse with photorealistic rendering. Due to the heterogeneously generated data types, foreground communication between Alice and David is infeasible. To fix this issue, we can apply the aforementioned 1)-4) steps so as to make Alice produce SRs that are compatible with David's~SM.

\section{Signaling Games for SMC} \label{Sec:SignalingGame}

\subsection{Lewis Signaling Games}


Game theory provides a mathematical modeling of agent strategic interactions. As a special case, signaling games focus on how meaningful signals or languages emerge through agent interactions. The Lewis signaling is the first of its kind \cite{Choi22_VTC}, which is widely used in biology and economics. To illustrate this, suppose Alice and Bob sequentially make decisions on sending and responding to signals, respectively. Alice has a set of states or \emph{types} that she can observe and understand (e.g., a type for the presence or absence of a predator when Alice is a mother bird and Bob is a newborn baby bird). For a given type, Alice aims to inform Bob by sending a signal, but Bob may not understand its meaning. To resolve this problem, Bob can build a signal-type mapping rule by training with rewards, and so does Alice for a signal-response mapping rule. These mapping rules are iteratively trained until each agent cannot find any better solution, i.e., until reaching the Nash equilibrium (NE).




\addtocounter{footnote}{-1}
\footnotetext{The 3D objects were generated by Magic3D \cite{https://doi.org/10.48550/arxiv.2211.10440}, a text-to-3D generative model, using the text prompts ``a ripe strawberry" and ``Sydney opera house, aerial view."}

\begin{figure}[t]
\centering
\includegraphics[width=.95\columnwidth]{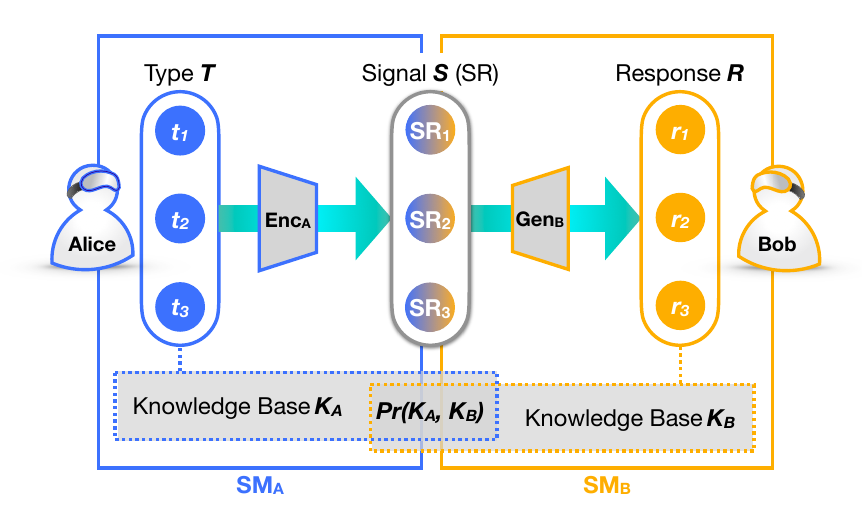} 
\caption{A signaling game-theoretic formulation of SMC with two agents, where Alice and Bob aim to learn the type-signal and signal-response mapping rules, respectively, while exploiting their local knowlwedge bases $K_A$ and $K_B$.} \label{Fig:4}
\end{figure}

We can interpret SMC through the lens of the Lewis game, as visualized in Fig.~\ref{Fig:4}. The type-signal and signal-response mapping rules can be seen as the semantic encoder and semantic generator in SMC, respectively. Accordingly, types, signals, and responses in the Lewis game coincide with the input multi-modal data, SRs, and generated data in SMC. Next, with more signals than types and responses, each type and response can be mapped into at least one unique signal. However, a variety of types and responses in the metaverse entail non-unique signal mappings. For instance, a generator may render the signal `bat' into a ``baseball bat" or ``animal bat" with uncertainty. This warrants the need for knowledge bases in SMC, to achieve intended responses in the Lewis game even with insufficient number of signals \cite{Choi22_VTC}. From this signaling game-theoretic point of view, we revisit the Levels A-C of SMC in Sec.~\ref{Sec:Intro}, and clarify major sources of errors in SMC as follows.
\begin{itemize}
\item \textbf{Level A -- Signal Distortion}: Erroneous receptions due to channel noise or interference would cause signaling errors. The distorted SRs due to DMC noise in Sec.~\ref{Sec:FLSL} fall in this category. This type of errors can be well quantified and analyzed via information theory. These errors can be mitigated by forward error correction or retransmissions after payload reduction, both of which can be enabled by exploiting the signal mapping rules.

\item \textbf{Level B -- Mapping Uncertainty}: Lack of signals and/or training in parallel (rather than sequentially) would cause a (partial) pooling NE wherein a signal is mapped into more than one type/response. In SMC, this happens when the number of available SRs is limited (e.g., due to small semantic encoder/decoder's output/input dimensions), or when some SRs are clustered for abstraction as we shall discuss in Sec.~\ref{Sec:NS_Transform}. The so-called pooling NE incurs erroneous rendering with uncertainty, which can be mitigated by utilizing KBs as side information~\cite{Choi22_VTC}.


\item \textbf{Level C -- Knowledge Inconsistency}: Consistent KBs can be used as additional information sources to reduce the mapping uncertainty. However, inconsistent KBs become not only useless, but also incur misinterpretation, necessitating background communication. In SMC, heterogeneous SMs studied in Sec.~\ref{Sec:DistributedML} fall in this category, wherein distributed learning as background communication is applied to make separately trained semantic encoder/generator NNs interoperable.
\end{itemize}

In fact, these three-level problems coincide with those in SC \cite{9955525,https://doi.org/10.48550/arxiv.2108.05681,Choi22_VTC,CLP_22,Guler18} following the Shannon-Weaver's communication hierarchy \cite{ShannonWeaver:1949}; namely, syntactic (i.e., bit-level), semantic, and pragmatic (or contextual) errors. This signaling game-theoretic interpretation facilitates a better understanding of the SC problems. Given distinct problem sources, different level problems inherently require different solutions. It also reveals that high-level solutions can assist in solving their lower level problems (e.g., joint syntatic and semantic communication design as a syntactic problem solution~\cite{CLP_22}), but the reverse is not true. This inter-level relationship also implies that high-level issues can even aggravate their lower level problems, emphasizing the importance of background communication as a highest-level solution.

\subsection{Dynamic Signaling Games}

The Lewis signaling game can be extended to incorporate more complex interactions and realistic environments. First, KBs can be external to agents (e.g., edge/cloud servers), and even become intelligent. Treating such an external intelligent KB as another agent in the game, this knowledgeable agent can be either a friend providing useful side information or a foe who intentionally shares malicious information, which is studied as the semantic communication game in \cite{Guler18}. Second, knowledge can be utilized not only for reducing the epistemic uncertainty of signal meanings, but also for signal compression. For the latter purpose, the rational speech act (RSA) model is relevant \cite{https://doi.org/10.48550/arxiv.2108.05681}. In the RSA model, all agents are assumed to be intelligent and in pursuit of signal minimality. For instance, if Alice knows that Bob is at a crossroad, then to make Bob `stop,' Alice only sends a shorter signal ``red" that can be understood as intended by Bob who believe that Alice already knows Bob's current situation.

Signaling games can play a key role in providing design guidelines to deep learning based frameworks. For instance, the vulnerability of DeepSC to heterogeneous agents discussed in Sec.~\ref{Sec:DeepSC} can be explained by the Level-B type error where mappings are given by trained transceivers. The need for exploiting its higher-level (Level-C) advocates the proposed solution in Sec.~\ref{Sec:FLSL}, which utilizes distributed learning as background communication. Notwithstanding, one fundamental limitation of signaling game-theoretic frameworks is due to their inherently fixed environments where types and signals are pre-determined and known to all agents. To complement such a drawback, for unknown types, it is possible to utilize NNs that can identify common patterns among many input samples (e.g., a simple NN can accurately classify the MNIST dataset comprising 60k samples that are all different). For time-varying types, MARL is effective in capturing response-type interactions, and the resultant temporal dynamics of types, which will be discussed in Sec.~\ref{Sec:MARL}.

\begin{figure}[t]
\centering
\includegraphics[width=\columnwidth]{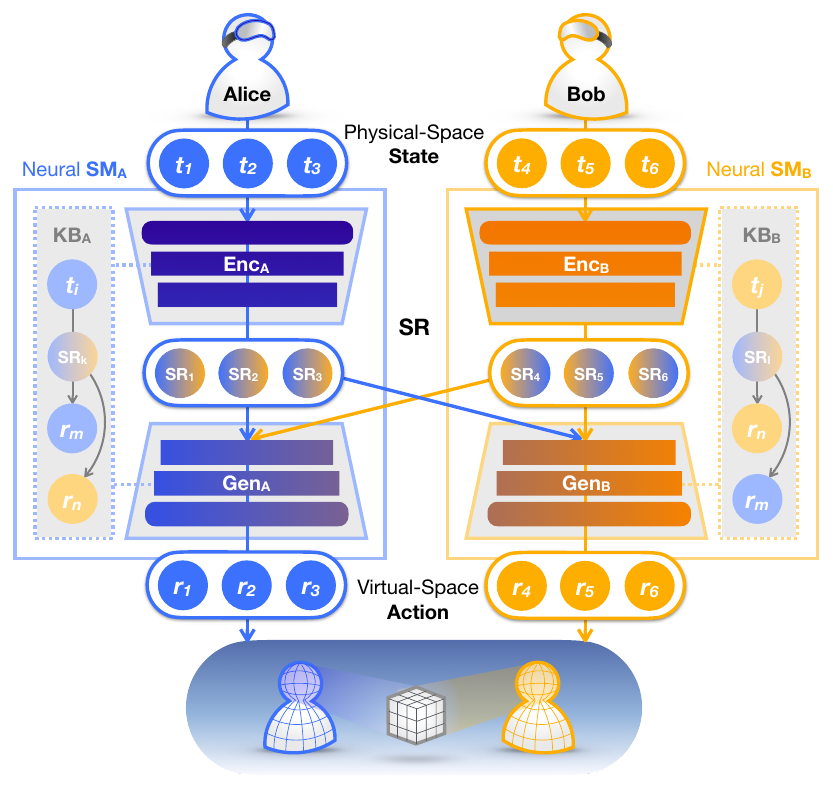} 
\caption{An MARL based formulation of SMC with two agents, where each agent's encoder-generator SM is an NN, and the NN's input-layer, hidden-layer, and output-layer activations correspond to the agent's states (types), SRs (signals), and actions (responses), respectively.} \label{Fig:5}
\end{figure}

\section{MARL and Symbolic AI for SMC} \label{Sec:MARL}

\subsection{MARL for Emergent Communication} \label{Sec:Comm-MARL}



Game theory can characterize optimal agents' state-action policies in a fixed environment, while reinforcement learning (RL) focuses on not only policy but also reflecting the environment that interacts with agents. Following this, signaling games assume fixed knowledge bases, as well as the types contained therein and the SRs of the types. With such fixed types and SRs, signaling games focus only on the emergent mapping between SRs and types. On the other hand, MARL can incorporate KBs updated over time. In this setting, agents' taking actions change environmental states, followed by the types in the knowledge bases. In this case, SRs should no longer be fixed. \emph{MARL with communication (Comm-MARL)} frameworks tackle this scenario \cite{https://doi.org/10.48550/arxiv.2203.08975}, and aim to learn not only the emergent mapping but also the emergent SRs. 

Precisely, Comm-MARL methods rest commonly on the centralized training with decentralize execution (CTDE) framework. In CTDE, a single NN is divided into a common critic NN shared by multiple actor NNs. Each actor NN implies an agent, and its input and output layer activations correspond to its environment observations and actions, respectively. During training, the critic valuates the actors' actions, and gives global rewards for training the entire NN. After training completes, the critic NN is lifted, and each actor NN becomes a {neural SM} integrating the semantic encoder/decoder and KB functionalities as discussed in Sec.~\ref{Sec:Architecture}. During and after the training process, several Comm-MARL frameworks additionally communicate each actor NN's hidden-layer activations, i.e., the SRs of their observations. Differentiable inter agent learning (DIAL), multi-agent deep deterministic policy gradient (MADDPG) \cite{https://doi.org/10.48550/arxiv.2203.08975}, and the modified distributed deep-Q learning \cite{seo2022towards} fall in this category. In such a Comm-MARL framework, at the beginning when agents are poor at solving a given task, the exchanged activations are nothing but random values. However, as training progresses, these activations gradually become emergent SRs that effectively coordinate agents for solving their task. 

With two agents, Fig.~\ref{Fig:5} visualizes their trained NNs as neural SMs. Notably, Comm-MARL can be flexibly extended to produce multi-level SRs by communicating multiple hidden activations at different layers (e.g., uplink SRs followed by downlink SRs at an upper layer as considered in~\cite{seo2022towards}).



\begin{figure}[t]
\centering
\includegraphics[width=.75\columnwidth]{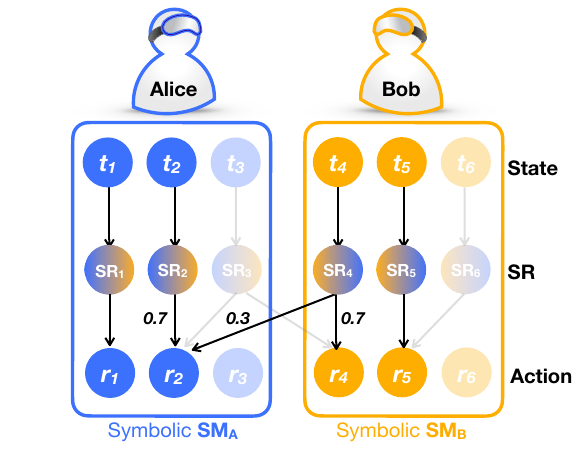} 
\caption{An example of neuro-symbolic transformation turning each agent's trained neural SM, i.e., encoder-generator NN, into a symbolic SM, i.e., a directed probabilistic graph, by treating the NN as a simulator, clustering equivalent activations and mapping patterns, and distinguishing uncertain patterns via their frequency of occurrence.} \label{Fig:6}
\end{figure}

\subsection{Neuro-Symbolic Transformation} \label{Sec:NS_Transform}

One succinct disadvantage of Comm-MARL is its difficulty in clarifying the SKs on SRs and their relationships that are stored in a black-box KB, i.e., an NN. To overcome this drawback, a neural SM can be transformed into its symbolic version \cite{seo2022towards}, i.e., \emph{symbolic SM}, as elaborated next.
\begin{enumerate}
    \item The entire set of trained actor NNs is a deterministic function having the agents' states as inputs as well as SRs and actions as outputs. Treating this function as a simulator, one can feed all possible state combinations into the function, and find the `state$\to$SR$\to$action' mappings that can be visualized as a directed symbolic graph.
    
    \item To make the symbolic graph concise and generalized via abstraction, redundant SRs (e.g., similar activations) and their mappings (e.g., different SRs yielding the same action) are clustered.

    \item For the SRs mapped to multiple agents, clustering may incur inter-agent interference (e.g., the same SR yielding different actions). To distinguish agent-specific messages, a probability is allocated to every edge of the graph. The probability is given by re-feeding all state combinations and counting the mapping frequency.
\end{enumerate}

Consequently, as Fig.~\ref{Fig:6} illustrates, the NN of Comm-MARL is transformed into a \emph{probabilistic symbolic graph} in the form of `state$\overset{}{\to}$SR$\overset{p}{\to}$action' which reads ``the SR yields a given action with probability $p$." In the symbolic graph, each edge with its connecting nodes can be separately expressed; for example as `SR$\overset{p}{\to}$action,' which is recast as `$\{p::\text{action}: -\text{SR} \}$.' according to the syntax of the \emph{probabilistic logic programming (ProbLog)} language in symbolic AI. Such symbolic expressions enable further manipulations such as Boolean algebra and reduction. Furthermore, \emph{semantic information theory (SIT)}~\cite{9955525} allows one to find the entropy of each symbolic expression using its associated probability $p$. Averaging such entropy, one can quantify the uncertainty or randomness of the entire symbolic graph (i.e., SM) or its subgraph, which can be useful for background SM communication. It is also remarkable that such a symbolic SM can be directly edited for coping with mission-critical factors (e.g., avoiding a certain action) and compatibility issues (e.g., replacing emergent SRs with pre-programmed signals).

\section{Conclusion and Future Directions} \label{Sec:Conclusion}

Inspired by generative AI and SC, in this article 
we proposed a novel framework of SMC for the wireless metaverse by decomposing the metaverse into distributed SMs. Enabling synchronous SMs relies on co-designing foreground SR communication and background SM communication under agent-specific heterogeneous SMs. To this end, we introduced promising modeling and algorithmic approaches based on distributed learning, Comm-MARL, signaling games, and symbolic AI.
To foster this new direction of SMC research, we conclude this article by identifying several open issues for future study.

\begin{itemize}
    \item \textbf{SM Consistency Measure}: Synchronizing SMs by reducing their element-wise differences, i.e., Hamming distance, may incur costly background communication, calling for a new SM consistency measure. In causal inference, the structured interventional distance quantifies the dissimilarity between two causal graphs by counting the output differences after taking a common intervention on both graphs. Similarly, in knowledge distillation of deep learning, a distillation regularizer measures the  the output difference between two NNs for a common input. Leveraging these approaches, SM consistency can be measured by counting the SM's output differences for the interactions within the scope of a given task.



    \item \textbf{Foreground-Background Joint Resource Allocation}:
    Foreground and background communications share common physical communication channels. Their joint communication resource allocation is non-trivial due not only to the closed-loop connection between foreground and background communiations, but also to the causal and logical relationships between SRs and SKs. In this perspective, learning causal SRs and constructing causal graphs of SRs as SKs are promising, which strengthen the causal impacts of SRs on SKs and clarify their relationships.



    \item \textbf{Knowledge Coordinating Infrastructure}: Communication infrastructure can play an important role in supporting SMC. For instance, an edge server enables transfering SMs to new agents based on task similarities could be resource-efficient, rather than constructing SMs from scratch. It can also translate SRs and transform SMs for backward compatibility with legacy communication and rendering architectures (e.g., a pre-programmed central metaverse server).

    
\end{itemize}

\bibliographystyle{ieeetr}
\bibliography{main}

\end{document}